\documentclass[sigconf]{acmart}
\usepackage{graphicx} \usepackage{enumitem}
\usepackage{pgf}
\usepackage{appendix}

\title{Examining~the~Effects~of~Degree~Distribution~and~Homophily in~Graph~Learning~Models}
\author{Mustafa Yasir}
\affiliation{\institution{Warwick University}
  \country{UK}}
\author{John Palowitch}
\affiliation{\institution{Google Research}
  \country{USA}}
\author{Anton Tsitsulin}
\affiliation{\institution{Google Research}
  \country{USA}}
\author{Long Tran-Thanh}
\affiliation{\institution{Warwick University}
  \country{UK}}
\author{Bryan Perozzi}
\affiliation{\institution{Google Research}
  \country{USA}}
\acmConference[KDD ADS '23]{KDD 2023 Conference Applied Data Science}{August 06--15,2023}{Long Beach, NY}

\begin{abstract}
Despite a surge in interest in GNN development, homogeneity in benchmarking datasets still presents a fundamental issue to GNN research. GraphWorld is a recent solution which uses the Stochastic Block Model (SBM) to generate diverse populations of synthetic graphs for benchmarking any GNN task. Despite its success, the SBM imposed fundamental limitations on the kinds of graph structure GraphWorld could create.  

In this work we examine how two additional synthetic graph generators can improve GraphWorld's evaluation; LFR, a well-established model in the graph clustering literature and CABAM, a recent adaptation of the Barabasi-Albert model tailored for GNN benchmarking. By integrating these generators, we significantly expand the coverage of graph space within the GraphWorld framework while preserving key graph properties observed in real-world networks. To demonstrate their effectiveness, we generate 300,000 graphs to benchmark 11 GNN models on a node classification task.  We find GNN performance variations in response to homophily, degree distribution and feature signal. Based on these findings, we classify models by their sensitivity to the new generators under these properties. Additionally, we release the extensions made to GraphWorld on the GitHub repository, offering further evaluation of GNN performance on new graphs.
\end{abstract}

\begin{document}
\maketitle
\keywords{graph neural networks, graph learning benchmarks, synthetic graphs}

\section{Introduction}
Interest in Graph Neural Networks (GNNs) has surged over the last decade, with thousands of new GNN variants being introduced \cite{keramatfar2022graph}. Despite this, a disproportionately small number of datasets from limited domains \cite{shchur2018pitfalls, Palowitch_2022} are used to benchmark and evaluate new models. The lack of diversity in benchmarking datasets presents a major issue in evaluating the empirical performance of GNNs. 

One approach to the problem of dataset variety is through the use of synthetic graph generators. These generators can create diverse yet controllable graph datasets that cover extensive regions of the space of all possible graphs. By using synthetic graphs, the problem of overfitting can be reduced, as each new dataset collection can be designed to be as diverse as possible. Furthermore, the field of Network Science offers a vast array of synthetic graph models \cite{barabasi2016network, watts1999swd} that can approximate important properties observed in real-world networks and serve as realistic benchmarks. This approach is employed by GraphWorld \cite{Palowitch_2022}, a novel methodology and system that relies on graph generation models to simulate statistically diverse graphs of arbitrary sizes for benchmarking purposes.

Currently, GraphWorld relies exclusively on the SBM to generate synthetic graphs. Despite its usefulness, evaluating GNN performance using results from a single graph generation model poses a limitation that goes against the challenges GraphWorld aims to address. 

In this work, we address this limitation by integrating two additional graph generation models into the system. We bring an established benchmark model with strong clustering properties: LFR, and a class-assortative adaptation of the Barabasi-Albert model: CABAM. By incorporating LFR and CABAM, we aim to (i) approximate properties observed in real-world networks that the SBM cannot replicate, and (ii) empirically expand the regions of graph space that GNNs can be benchmarked on.

Additionally, we conduct node classification experiments using 11 GNN models. We reveal variations in GNN performance in response to homophily, degree distribution and feature signal across graph generators. Through these results, we provide a classification of GNN models with respect to their sensitivity to the newly introduced graph generators. Furthermore, we make the extensions to GraphWorld publicly available in aid of further analysis on GNN performance.

We summarise our contributions as follows:
\begin{itemize}[topsep=3pt]
    \item \textbf{Extending benchmark datasets}. We integrate two additional graph generators with GraphWorld that  (i) exhibit desirable properties observed in real-world networks and (ii) cover new regions of graph space for GNN benchmarking.
    \item \textbf{Performance classifications}. We run benchmarking experiments with 11 GNN models and find performance variations in response to homophily, degree distribution and feature signal. We classify models by their sensitivity to the new generators under these properties.
    \item \textbf{Code}. We release the extensions made to GraphWorld on the GitHub repository.\footnote{Our contributions are available at \url{https://github.com/google-research/graphworld}}
\end{itemize}

\section{Graph Generation Models}

For additional graph generation models, we look to the field of Network Science; where a large number of models have been introduced in the aim of producing synthetic graphs that exhibit properties observed in real world networks. The properties in focus being: degree distribution, community structure and homophily. In this section, we detail the relative strengths of each model with respect to these properties.

\subsection{Stochastic Block Model}
The SBM serves as the original graph model used in GraphWorld, designed to generate graphs that exhibit communities: densely connected and well-separated subsets of nodes. This is accomplished by introducing a parameterized community distribution and a matrix of edge probabilities. The graph generation process begins by partitioning the node set based on the community distribution. Then, edges are assigned randomly to nodes within and between communities, following the edge probability matrix.

The popularity of the SBM in benchmarking community detection algorithms has resulted in the development of several variants aimed at more accurately modelling real-world networks. One such variant, used in GraphWorld, is the Degree Corrected Stochastic Block Model (DC-SBM) \cite{Karrer_2011}. The DC-SBM incorporates heterogeneity in vertex degrees, making it possible to replicate arbitrary degree sequences within the model. In GraphWorld, we leverage this capability by generating artificial degree sequences that follow arbitrary power law exponents. This allows us to generate SBM graphs with user-defined ranges of degree distributions.

However, the SBM relies on fixed values defined in its edge probability matrix, making it unable to precisely replicate a given degree sequence and generate graphs that adhere to a true power law. This limitation motivates the choice of LFR and CABAM as complimentary models. Both LFR and CABAM utilize generative processes to create graphs with natural power law degree distributions.

\subsection{CABAM}
CABAM generates \textbf{C}lass-\textbf{A}ssortative graphs via the \textbf{B}arabasi-\textbf{A}lbert (BA) \textbf{M}odel. The BA model represents a significant improvement over traditional graph models by approximating real-world graph properties through two key concepts of (i) growth and (ii) preferential attachment \cite{barabasi2016network}. These account for the observations that (i) real-world networks emerge through the continual addition of nodes and (ii) newly added nodes exhibit a preference for connecting to highly connected nodes.

Following this generative process, the BA model generates scale-free networks, which are characterized by degree distributions that conform to a true power law. This forms a fundamental property observed in numerous real-world networks \cite{Clauset_2009}. CABAM extends the BA model by introducing parameterized community generation, whilst preserving the original scale-free property. In GraphWorld, these parameters are used to precisely control the community distribution of a CABAM graph.

Moreover, CABAM offers the flexibility to determine edge homophily in various ways: it can be constant, dependent on node degree, or follow any arbitrary function. In GraphWorld, we preserve this feature and employ similar methods as described for the SBM to create arbitrarily-diverse graph structures under these properties. Indeed, parameterizing these properties fulfills the requirements of synthetic graph benchmarking for GNNs. It is widely observed that community distribution and homophily exhibit significant variations across real-world networks \cite{cabam2020shah}, and these variations can lead to substantial performance differences among GNNs \cite{Suresh_2021}.

The strength of CABAM lies in its ability to generate BA-style graphs with tune-able homophily and community structure. However, the degree distributions of CABAM graphs are predetermined by the described generative process, resulting in a fixed power law distribution with an exponent of $-3$. Considering this limitation, we seek an additional model that can generate scale-free graphs parameterized in both community structure and degree distribution. 

\begin{figure}[t]
\centering
\includegraphics[width=0.6\linewidth]{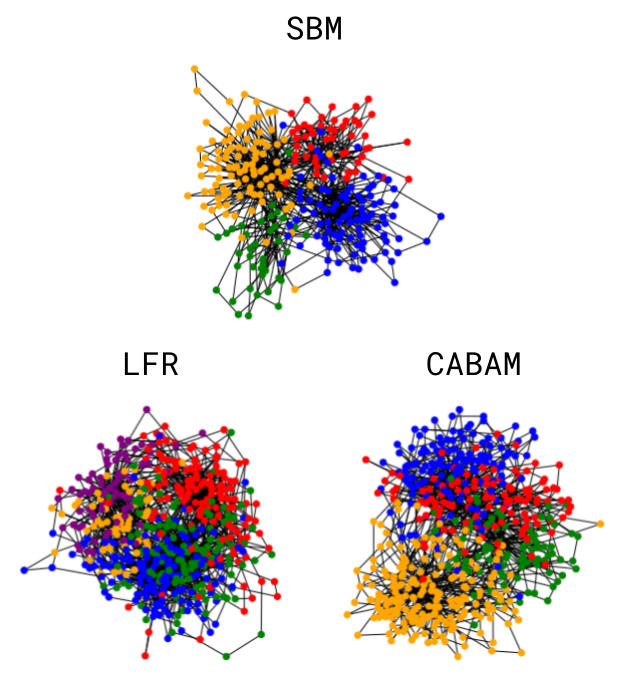}
\caption{\label{fig:network-plots}NetworkX plots of 512-node graphs from SBM and the two new node classification dataset generators CABAM and LFR. We align input parameters to the models, showing similarities in some high-level properties such as number of communities, whereas there are differences in low-level properties such as degree distribution and triangle count.}
\end{figure}

\subsection{LFR}
LFR refers to the \textbf{L}ancichinetti–\textbf{F}ortunato–\textbf{R}adicchi (LFR) benchmark \cite{Lancichinetti_2008}. The LFR benchmark extends synthetic benchmarking graphs by introducing heterogeneity in both: node degree, similar to the DC-SBM, and community size. In particular, it allows for the creation of graphs whose degree and community size distributions follow power laws with distinct, arbitrary exponents. With the inclusion of a mixing parameter that determines edge homophily, LFR provides a complimentary resource to study the effects of homophily, degree distribution and community structure on GNN performance.

The inclusion of heterogeneity in both node degree and community size addresses the need to better approximate real-world networks, where communities and degrees can exhibit arbitrary, non-uniform sizes. Since these properties are also known to influence GNN performance \cite{Hussain_2021}, GraphWorld exposes community and degree distribution parameters to the user, allowing for the generation of LFR benchmark datasets following arbitrary, randomly sampled power law distributions. To further capture the unique community structure, we introduce a feature that allows SBM and CABAM graphs to replicate the community structure of a given LFR graph on the same node count. 

In addition to its graph properties, LFR's well-established popularity as a benchmark for community detection algorithms provides further motivation. By introducing LFR to the benchmarking of modern graph learning algorithms like GNNs, where this benchmark is unseen, we provide a novel addition to the array of synthetic graph datasets for GNN benchmarking.

\section{Experiments}
With the unique properties of each model established, in this section we run novel GraphWorld experiments using these models, with the aim of answering the following research questions:
\begin{enumerate}
    \item Can LFR and CABAM allow GraphWorld to produce new regions of graph space beyond the SBM?
    \item If so, do the new regions produced by LFR and CABAM show new differentials between groups of GNN models?
    \item What new insights about GNN models can be learned from the regions of graph space produced by LFR and CABAM?
\end{enumerate}

\subsection{Experimental Design}
Using GraphWorld, we generate 100,000 graph samples from each generator described in the previous section (SBM, CABAM and LFR). Whilst we discussed a subset of the parameters for each generator earlier, in Appendix A, we provide a full description of all the parameters for each generator along with their respective input values. These tables provide enough information for the same GraphWorld experiments to be readily reproduced.

In order to study the global effects of graph properties such as homophily and degree distribution on GNN performance, we match the inputs of parameters that perform similar functions across generators. This is possible due to the selection of graph models that share similar controls over such properties. For example, following the discussion on the parameterization of degree distribution on SBM and LFR graphs, we ensure that graph samples across all generators receive random values from similar ranges in their respective degree sequence generation parameters. By matching similar parameters across generators, we ensure a fair \textit{global} comparison of GNN performance across different graph generators.

Furthermore, by varying parameter values across wide ranges, we can conduct \textit{local} analyses to examine the specific effects of individual parameters within each generator on GNN performance. GraphWorld computes metrics of each sampled graph to then quantify the effects of such parameters in a unified manner to answer research questions 1 and 2. For example, whilst the \textit{control} of homophily through input parameters may vary across generators, we use the edge homogeneity statistic provided in GraphWorld's output to \textit{measure} the homophily of each graph after the graph generation process.

Following the experimental framework outlined in the GraphWorld paper \cite{Palowitch_2022}, we use the generated graph samples to benchmark 11 GNN models: \textbf{ARMA} \cite{Bianchi_2021}, \textbf{APPNP} \cite{gasteiger2022predict}, \textbf{FiLM} \cite{brockschmidt2020gnnfilm}, \textbf{GAT} \cite{veličković2018graph}, \textbf{GATv2} \cite{brody2022attentive}, \textbf{GCN} \cite{kipf2017semisupervised}, \textbf{GIN} \cite{xu2019powerful}, \textbf{GraphSAGE} \cite{hamilton2018inductive}, \textbf{SGC} \cite{wu2019simplifying}, \textbf{SuperGAT} \cite{kim2022friendly}, \textbf{Transformer} \cite{shi2021masked} and 2 baselines: \textbf{Multi-Layer Perceptron}, \textbf{Personalized PageRank} \cite{334}, on a node classification task. 

\subsection{Properties of  Graph Statistics}
To answer research question 1 we compute kernel density estimate plots comparing the generated graph samples on 6 metrics provided in GraphWorld's output. We use the metrics described below to study the distribution of graphs on the following properties:
\begin{itemize}[topsep=3pt]
    \item \textbf{Degree Distribution:} Power law estimate and degree gini coefficient
    \item \textbf{Homophily:} Edge homogeneity
    \item \textbf{Community Structure:} Average clustering coefficient (avg\_cc), simpsons community size and the number of triangles
\end{itemize}

\begin{figure}[t]
  \centering
  \includegraphics[width=\columnwidth]{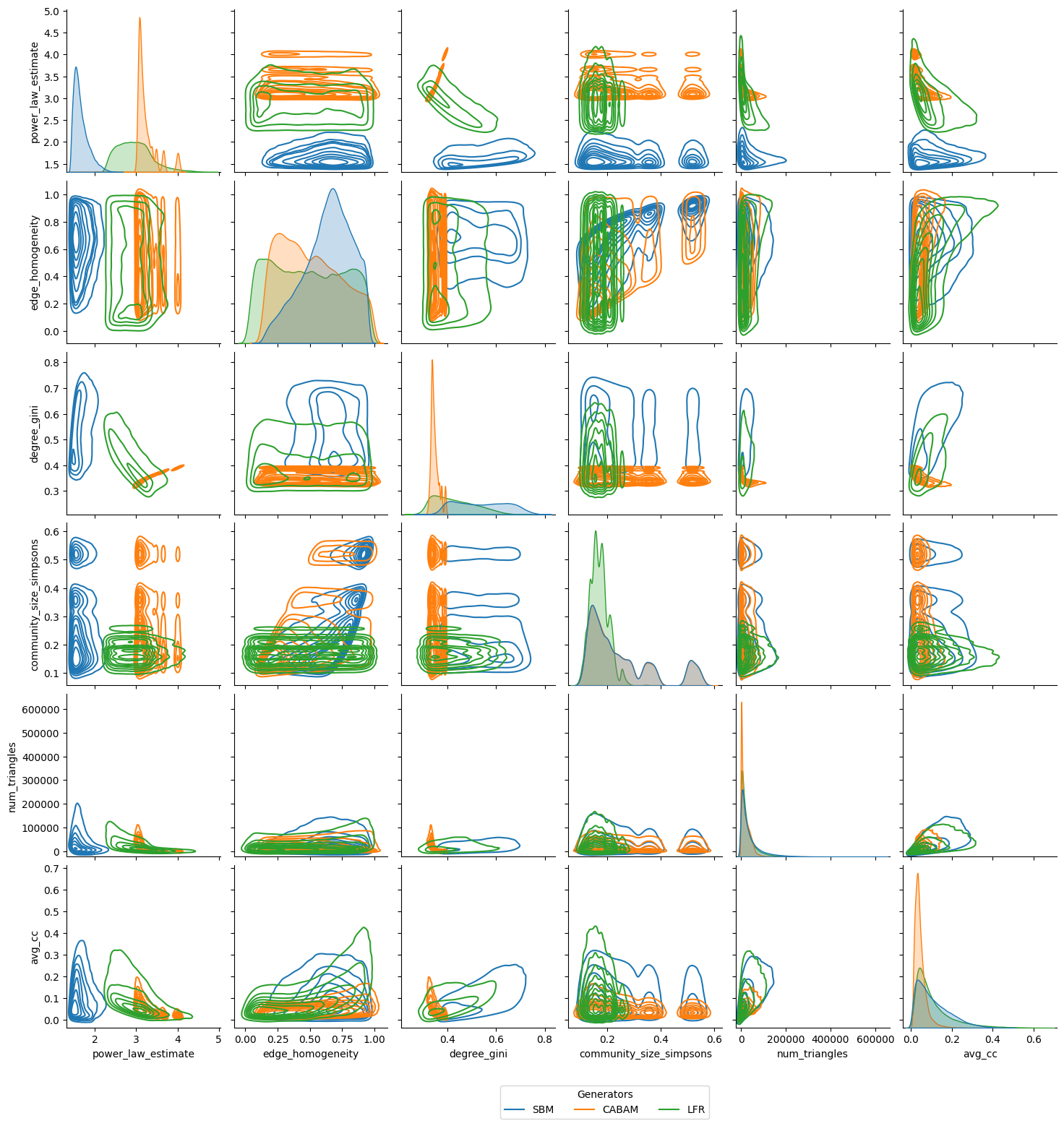}
  \caption{Kernel Density Estimates of SBM, CABAM and LFR graphs on various metrics. Our results show that CABAM and LFR graphs exist in non-overlapping regions of graph space when compared to the SBM on metrics measuring degree sequence and homophily.\label{fig:graph-stats}}
\end{figure}

\paragraph{Insights.} 
Our results in Figure~\ref{fig:graph-stats} show that CABAM and LFR indeed cover new regions of graph space beyond the SBM. Specifically, CABAM and LFR graphs exhibit degree distributions vastly different from the SBM. For CABAM this is characterized by the large non-overlapping spikes in its power law estimate distributions, whilst for LFR we see a much wider range of power laws compared to either generator.

Additionally, the edge homogeneity distributions show that CABAM and LFR graphs exhibit a wider range of homophily over SBM graphs, particularly in the lower homophily regions. In summary, our results show that CABAM and LFR widen the scope of graphs we can generate by introducing unique degree distributions and wider ranges of homophily. 

\subsection{GNN Benchmarking Results}
To answer research questions 2 and 3, we look at the performance of GNNs on each of the 300,000 graph samples from the GraphWorld benchmarking experiments. We measure the performance of a given model with the ROC-AUC-One-Vs-Rest score and plot this against the edge homogeneity, power law estimate and feature center distance (feature signal) of each graph. Thus for each GNN model we have 3 curves, representing the performance on SBM, CABAM and LFR graphs. 

\begin{figure}[t]
  \centering
  \includegraphics[width=\columnwidth]{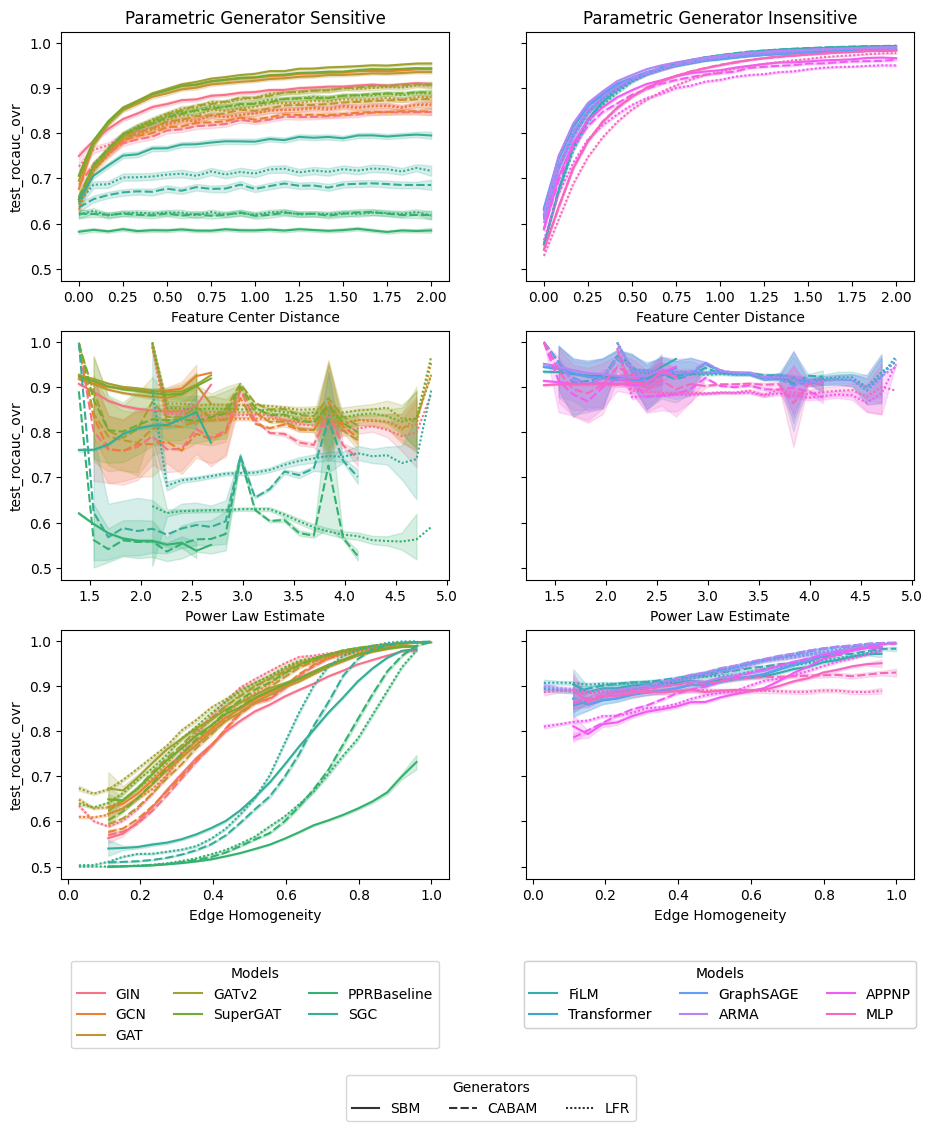}
  \caption{ROC-AUC of models in GraphWorld.  Our results show that GNN models form two groups with respect to parametric graph generators: sensitive (GIN, GCN, GAT, GATv2, SuperGAT, SGC), and insensitive (FiLM, Transformer, GraphSAGE, ARMA and APPNP).\label{fig:gnn-variation}}
\end{figure}

In Figure~\ref{fig:gnn-variation}, we take the performance curves of each GNN model and group them into 2 columns, one for each group of GNN models exhibiting similar performance. We have each row corresponding to a graph metric and each column corresponding to a group of GNN models.

\paragraph{Insights.}
In Figure~\ref{fig:gnn-variation}, our results demonstrate the ability to classify GNN models based on their sensitivity to the generators used for benchmarking. We observe that certain groups of GNN models exhibit variations in their performance across SBM, CABAM, and LFR graphs, indicating their sensitivity to the specific properties introduced by these generators. These models fall into the parametric generator-sensitive group, reflecting the larger variation in their performance across different graph models. This variation arises due to the inclusion of LFR and CABAM graphs, which cover previously unexplored regions of the graph space.

However, our analysis also reveals the existence of a parametric generator-insensitive group of GNN models. These models demonstrate minimal variation in their performance across the SBM, CABAM, and LFR graphs, suggesting that they are robust to the specific characteristics introduced by these generators. Despite the introduction of new graph models, these GNN models exhibit consistent and stable performance, highlighting their insensitivity to the variations in graph structure generated by the different models.

The observed behavior can be attributed to the characteristics of the GNN models within the parametric generator-insensitive group. These models tend to be less sensitive to the local graph structure and instead focus on more global information. For example, models such as APPNP, GraphSAGE, and Transformer employ smoothing techniques, which allow them to mitigate the influence of specific graph structures and instead capture more general patterns in the data.

Furthermore, the parametric generator-insensitive models exhibit a higher degree of reliance on node features. This is evident from the edge homogeneity plots, which indicate that these models achieve significantly better performance than the sensitive group on graphs with lower levels of homophily. This suggests that these models are more adept at leveraging the available node features to make accurate predictions. However, it is worth noting that these models also experience a larger drop in performance on graphs with low feature signal, indicating that the quality and informativeness of the features play a crucial role in their performance.

In summary, our findings highlight the presence of new differentials between groups of GNN models when exploring the additional regions of graph space covered by LFR and CABAM.

\section{Conclusion}
In this work, we examined how two additional synthetic graph generators, LFR and CABAM, could improve the evaluation of GNN performance in GraphWorld. We established the unique properties of each generator and generated 300,000 graph samples which showed that the newly introduced models cover new regions of graph space beyond the SBM. We used these samples to benchmark 11 GNN models and found variations in the sensitivity of GNN models in response to the new graph generators. We classified models as being either parametric generator-sensitive or parametric generator-insensitive. The parametric generator-insensitive models, characterized by their focus on global information and reliance on node features, exhibit distinct performance patterns compared to the sensitive group. Understanding these differences provides valuable insights into the behavior and strengths of different GNN models across a range of graph structures. Our work addresses the critical issue of dataset homogeneity in GNN research and offers an advancement in generating diverse and realistic synthetic benchmarks. By making the extensions to GraphWorld publicly available and providing the parameter sets used to conduct these experiments, we open doors for further investigation into the factors affecting GNN performance.

\bibliographystyle{ACM-Reference-Format}
\bibliography{bibliography}


\begin{thebibliography}{23}


\ifx \showCODEN    \undefined \def \showCODEN     #1{\unskip}     \fi
\ifx \showDOI      \undefined \def \showDOI       #1{#1}\fi
\ifx \showISBNx    \undefined \def \showISBNx     #1{\unskip}     \fi
\ifx \showISBNxiii \undefined \def \showISBNxiii  #1{\unskip}     \fi
\ifx \showISSN     \undefined \def \showISSN      #1{\unskip}     \fi
\ifx \showLCCN     \undefined \def \showLCCN      #1{\unskip}     \fi
\ifx \shownote     \undefined \def \shownote      #1{#1}          \fi
\ifx \showarticletitle \undefined \def \showarticletitle #1{#1}   \fi
\ifx \showURL      \undefined \def \showURL       {\relax}        \fi
\providecommand\bibfield[2]{#2}
\providecommand\bibinfo[2]{#2}
\providecommand\natexlab[1]{#1}
\providecommand\showeprint[2][]{arXiv:#2}

\bibitem[Barabási and Pósfai(2016)]%
        {barabasi2016network}
\bibfield{author}{\bibinfo{person}{Albert-László Barabási} {and}
  \bibinfo{person}{Márton Pósfai}.} \bibinfo{year}{2016}\natexlab{}.
\newblock \bibinfo{booktitle}{\emph{Network science}}.
\newblock \bibinfo{publisher}{Cambridge University Press},
  \bibinfo{address}{Cambridge}.
\newblock
\showISBNx{9781107076266 1107076269}
\urldef\tempurl%
\url{http://barabasi.com/networksciencebook/}
\showURL{%
\tempurl}


\bibitem[Bianchi et~al\mbox{.}(2021)]%
        {Bianchi_2021}
\bibfield{author}{\bibinfo{person}{Filippo~Maria Bianchi},
  \bibinfo{person}{Daniele Grattarola}, \bibinfo{person}{Lorenzo Livi}, {and}
  \bibinfo{person}{Cesare Alippi}.} \bibinfo{year}{2021}\natexlab{}.
\newblock \showarticletitle{Graph Neural Networks with Convolutional {ARMA}
  Filters}.
\newblock \bibinfo{journal}{\emph{{IEEE} Transactions on Pattern Analysis and
  Machine Intelligence}} (\bibinfo{year}{2021}), \bibinfo{pages}{1--1}.
\newblock
\urldef\tempurl%
\url{https://doi.org/10.1109/tpami.2021.3054830}
\showDOI{\tempurl}


\bibitem[Brin and Page(1998)]%
        {334}
\bibfield{author}{\bibinfo{person}{Sergey Brin} {and} \bibinfo{person}{Lawrence
  Page}.} \bibinfo{year}{1998}\natexlab{}.
\newblock \showarticletitle{The Anatomy of a Large-Scale Hypertextual Web
  Search Engine}.
\newblock \bibinfo{journal}{\emph{Computer Networks}}  \bibinfo{volume}{30}
  (\bibinfo{year}{1998}), \bibinfo{pages}{107--117}.
\newblock
\urldef\tempurl%
\url{http://www-db.stanford.edu/~backrub/google.html}
\showURL{%
\tempurl}


\bibitem[Brockschmidt(2020)]%
        {brockschmidt2020gnnfilm}
\bibfield{author}{\bibinfo{person}{Marc Brockschmidt}.}
  \bibinfo{year}{2020}\natexlab{}.
\newblock \bibinfo{title}{GNN-FiLM: Graph Neural Networks with Feature-wise
  Linear Modulation}.
\newblock
\newblock
\showeprint[arxiv]{1906.12192}~[cs.LG]


\bibitem[Brody et~al\mbox{.}(2022)]%
        {brody2022attentive}
\bibfield{author}{\bibinfo{person}{Shaked Brody}, \bibinfo{person}{Uri Alon},
  {and} \bibinfo{person}{Eran Yahav}.} \bibinfo{year}{2022}\natexlab{}.
\newblock \bibinfo{title}{How Attentive are Graph Attention Networks?}
\newblock
\newblock
\showeprint[arxiv]{2105.14491}~[cs.LG]


\bibitem[Clauset et~al\mbox{.}(2009)]%
        {Clauset_2009}
\bibfield{author}{\bibinfo{person}{Aaron Clauset},
  \bibinfo{person}{Cosma~Rohilla Shalizi}, {and} \bibinfo{person}{M.~E.~J.
  Newman}.} \bibinfo{year}{2009}\natexlab{}.
\newblock \showarticletitle{Power-Law Distributions in Empirical Data}.
\newblock \bibinfo{journal}{\emph{SIAM Rev.}} \bibinfo{volume}{51},
  \bibinfo{number}{4} (\bibinfo{date}{nov} \bibinfo{year}{2009}),
  \bibinfo{pages}{661--703}.
\newblock
\urldef\tempurl%
\url{https://doi.org/10.1137/070710111}
\showDOI{\tempurl}


\bibitem[Gasteiger et~al\mbox{.}(2022)]%
        {gasteiger2022predict}
\bibfield{author}{\bibinfo{person}{Johannes Gasteiger},
  \bibinfo{person}{Aleksandar Bojchevski}, {and} \bibinfo{person}{Stephan
  Günnemann}.} \bibinfo{year}{2022}\natexlab{}.
\newblock \bibinfo{title}{Predict then Propagate: Graph Neural Networks meet
  Personalized PageRank}.
\newblock
\newblock
\showeprint[arxiv]{1810.05997}~[cs.LG]


\bibitem[Hamilton et~al\mbox{.}(2018)]%
        {hamilton2018inductive}
\bibfield{author}{\bibinfo{person}{William~L. Hamilton}, \bibinfo{person}{Rex
  Ying}, {and} \bibinfo{person}{Jure Leskovec}.}
  \bibinfo{year}{2018}\natexlab{}.
\newblock \bibinfo{title}{Inductive Representation Learning on Large Graphs}.
\newblock
\newblock
\showeprint[arxiv]{1706.02216}~[cs.SI]


\bibitem[Hussain et~al\mbox{.}(2021)]%
        {Hussain_2021}
\bibfield{author}{\bibinfo{person}{Hussain Hussain}, \bibinfo{person}{Tomislav
  Duricic}, \bibinfo{person}{Elisabeth Lex}, \bibinfo{person}{Roman Kern},
  {and} \bibinfo{person}{Denis Helic}.} \bibinfo{year}{2021}\natexlab{}.
\newblock \showarticletitle{On the Impact of Communities on Semi-supervised
  Classification Using Graph Neural Networks}.
\newblock In \bibinfo{booktitle}{\emph{Studies in Computational Intelligence}}.
  \bibinfo{publisher}{Springer International Publishing},
  \bibinfo{pages}{15--26}.
\newblock
\urldef\tempurl%
\url{https://doi.org/10.1007/978-3-030-65351-4_2}
\showDOI{\tempurl}


\bibitem[Karrer and Newman(2011)]%
        {Karrer_2011}
\bibfield{author}{\bibinfo{person}{Brian Karrer} {and}
  \bibinfo{person}{M.~E.~J. Newman}.} \bibinfo{year}{2011}\natexlab{}.
\newblock \showarticletitle{Stochastic blockmodels and community structure in
  networks}.
\newblock \bibinfo{journal}{\emph{Physical Review E}} \bibinfo{volume}{83},
  \bibinfo{number}{1} (\bibinfo{date}{jan} \bibinfo{year}{2011}).
\newblock
\urldef\tempurl%
\url{https://doi.org/10.1103/physreve.83.016107}
\showDOI{\tempurl}


\bibitem[Keramatfar et~al\mbox{.}(2022)]%
        {keramatfar2022graph}
\bibfield{author}{\bibinfo{person}{Abdalsamad Keramatfar},
  \bibinfo{person}{Mohadeseh Rafiee}, {and} \bibinfo{person}{Hossein
  Amirkhani}.} \bibinfo{year}{2022}\natexlab{}.
\newblock \bibinfo{title}{Graph Neural Networks: a bibliometrics overview}.
\newblock
\newblock
\showeprint[arxiv]{2201.01188}~[cs.LG]


\bibitem[Kim and Oh(2022)]%
        {kim2022friendly}
\bibfield{author}{\bibinfo{person}{Dongkwan Kim} {and} \bibinfo{person}{Alice
  Oh}.} \bibinfo{year}{2022}\natexlab{}.
\newblock \bibinfo{title}{How to Find Your Friendly Neighborhood: Graph
  Attention Design with Self-Supervision}.
\newblock
\newblock
\showeprint[arxiv]{2204.04879}~[cs.LG]


\bibitem[Kipf and Welling(2017)]%
        {kipf2017semisupervised}
\bibfield{author}{\bibinfo{person}{Thomas~N. Kipf} {and} \bibinfo{person}{Max
  Welling}.} \bibinfo{year}{2017}\natexlab{}.
\newblock \bibinfo{title}{Semi-Supervised Classification with Graph
  Convolutional Networks}.
\newblock
\newblock
\showeprint[arxiv]{1609.02907}~[cs.LG]


\bibitem[Lancichinetti et~al\mbox{.}(2008)]%
        {Lancichinetti_2008}
\bibfield{author}{\bibinfo{person}{Andrea Lancichinetti},
  \bibinfo{person}{Santo Fortunato}, {and} \bibinfo{person}{Filippo Radicchi}.}
  \bibinfo{year}{2008}\natexlab{}.
\newblock \showarticletitle{Benchmark graphs for testing community detection
  algorithms}.
\newblock \bibinfo{journal}{\emph{Physical Review E}} \bibinfo{volume}{78},
  \bibinfo{number}{4} (\bibinfo{date}{oct} \bibinfo{year}{2008}).
\newblock
\urldef\tempurl%
\url{https://doi.org/10.1103/physreve.78.046110}
\showDOI{\tempurl}


\bibitem[Palowitch et~al\mbox{.}(2022)]%
        {Palowitch_2022}
\bibfield{author}{\bibinfo{person}{John Palowitch}, \bibinfo{person}{Anton
  Tsitsulin}, \bibinfo{person}{Brandon Mayer}, {and} \bibinfo{person}{Bryan
  Perozzi}.} \bibinfo{year}{2022}\natexlab{}.
\newblock \showarticletitle{{GraphWorld}}. In
  \bibinfo{booktitle}{\emph{Proceedings of the 28th {ACM} {SIGKDD} Conference
  on Knowledge Discovery and Data Mining}}. \bibinfo{publisher}{{ACM}}.
\newblock
\urldef\tempurl%
\url{https://doi.org/10.1145/3534678.3539203}
\showDOI{\tempurl}


\bibitem[Shah(2020)]%
        {cabam2020shah}
\bibfield{author}{\bibinfo{person}{Neil Shah}.}
  \bibinfo{year}{2020}\natexlab{}.
\newblock \showarticletitle{Scale-Free, Attributed and Class-Assortative Graph
  Generation to Facilitate Introspection of Graph Neural Networks}. In
  \bibinfo{booktitle}{\emph{KDD Mining and Learning with Graphs}}.
\newblock


\bibitem[Shchur et~al\mbox{.}(2018)]%
        {shchur2018pitfalls}
\bibfield{author}{\bibinfo{person}{Oleksandr Shchur},
  \bibinfo{person}{Maximilian Mumme}, \bibinfo{person}{Aleksandar Bojchevski},
  {and} \bibinfo{person}{Stephan G{\"u}nnemann}.}
  \bibinfo{year}{2018}\natexlab{}.
\newblock \showarticletitle{Pitfalls of graph neural network evaluation}.
\newblock \bibinfo{journal}{\emph{arXiv preprint arXiv:1811.05868}}
  (\bibinfo{year}{2018}).
\newblock


\bibitem[Shi et~al\mbox{.}(2021)]%
        {shi2021masked}
\bibfield{author}{\bibinfo{person}{Yunsheng Shi}, \bibinfo{person}{Zhengjie
  Huang}, \bibinfo{person}{Shikun Feng}, \bibinfo{person}{Hui Zhong},
  \bibinfo{person}{Wenjin Wang}, {and} \bibinfo{person}{Yu Sun}.}
  \bibinfo{year}{2021}\natexlab{}.
\newblock \bibinfo{title}{Masked Label Prediction: Unified Message Passing
  Model for Semi-Supervised Classification}.
\newblock
\newblock
\showeprint[arxiv]{2009.03509}~[cs.LG]


\bibitem[Suresh et~al\mbox{.}(2021)]%
        {Suresh_2021}
\bibfield{author}{\bibinfo{person}{Susheel Suresh}, \bibinfo{person}{Vinith
  Budde}, \bibinfo{person}{Jennifer Neville}, \bibinfo{person}{Pan Li}, {and}
  \bibinfo{person}{Jianzhu Ma}.} \bibinfo{year}{2021}\natexlab{}.
\newblock \showarticletitle{Breaking the Limit of Graph Neural Networks by
  Improving the Assortativity of Graphs with Local Mixing Patterns}. In
  \bibinfo{booktitle}{\emph{Proceedings of the 27th {ACM} {SIGKDD} Conference
  on Knowledge Discovery {\&} Data Mining}}. \bibinfo{publisher}{{ACM}}.
\newblock
\urldef\tempurl%
\url{https://doi.org/10.1145/3447548.3467373}
\showDOI{\tempurl}


\bibitem[Veličković et~al\mbox{.}(2018)]%
        {veličković2018graph}
\bibfield{author}{\bibinfo{person}{Petar Veličković},
  \bibinfo{person}{Guillem Cucurull}, \bibinfo{person}{Arantxa Casanova},
  \bibinfo{person}{Adriana Romero}, \bibinfo{person}{Pietro Liò}, {and}
  \bibinfo{person}{Yoshua Bengio}.} \bibinfo{year}{2018}\natexlab{}.
\newblock \bibinfo{title}{Graph Attention Networks}.
\newblock
\newblock
\showeprint[arxiv]{1710.10903}~[stat.ML]


\bibitem[Watts(1999)]%
        {watts1999swd}
\bibfield{author}{\bibinfo{person}{D.J. Watts}.}
  \bibinfo{year}{1999}\natexlab{}.
\newblock \bibinfo{booktitle}{\emph{{Small Worlds: the dynamics of networks
  between order and randomness}}}.
\newblock \bibinfo{publisher}{Princeton Univ Pr}.
\newblock


\bibitem[Wu et~al\mbox{.}(2019)]%
        {wu2019simplifying}
\bibfield{author}{\bibinfo{person}{Felix Wu}, \bibinfo{person}{Tianyi Zhang},
  \bibinfo{person}{Amauri~Holanda de Souza Jr.~au2},
  \bibinfo{person}{Christopher Fifty}, \bibinfo{person}{Tao Yu}, {and}
  \bibinfo{person}{Kilian~Q. Weinberger}.} \bibinfo{year}{2019}\natexlab{}.
\newblock \bibinfo{title}{Simplifying Graph Convolutional Networks}.
\newblock
\newblock
\showeprint[arxiv]{1902.07153}~[cs.LG]


\bibitem[Xu et~al\mbox{.}(2019)]%
        {xu2019powerful}
\bibfield{author}{\bibinfo{person}{Keyulu Xu}, \bibinfo{person}{Weihua Hu},
  \bibinfo{person}{Jure Leskovec}, {and} \bibinfo{person}{Stefanie Jegelka}.}
  \bibinfo{year}{2019}\natexlab{}.
\newblock \bibinfo{title}{How Powerful are Graph Neural Networks?}
\newblock
\newblock
\showeprint[arxiv]{1810.00826}~[cs.LG]


\end{thebibliography}

\appendix

\section{Generator Parameters}
\begin{table*}
  \caption{SBM}
  \label{tab:SBM}
  \begin{tabular}{lll}
    \toprule
    Parameter Name & Description & Values\\
    \midrule
    nvertex & Number of vertices & [1028,4096] \\
    avg. degree & Average expected node degree & [1,32]\\
    min degree & Minimum degree & [2,20] \\
    $p/q$ ratio &  Ratio of intra-community to inter-community edge probabilities & [1,16]\\
    exponent & Value of the power law exponent used to generate expected node degrees & [0.2,3.0]\\
    num clusters & Number of communities & [2,10]\\
    cluster size slope & Slope of cluster sizes when index-ordered by size & [0.0,1.0]\\
    feature center distance & Variance of feature cluster centers, generated from a multivariate Normal & [0.0,2.0]\\
    feature dim & Dimension of feature vector assigned to every node & [16,16]\\
  \bottomrule
\end{tabular}
\end{table*}

\begin{table*}
  \caption{CABAM}
  \label{tab:CABAM}
  \begin{tabular}{lll}
    \toprule
    Parameter Name & Description & Values\\
    \midrule
    nvertex & Number of vertices & [1028,4096] \\
    min degree & Minimum degree & [2,20] \\
    inter link strength &  Probability of a node forming an intra-community edge & [0.5,1]\\
    num clusters & Number of communities & [2,10]\\
    cluster size slope & Slope of cluster sizes when index-ordered by size & [0.0,1.0]\\
    feature center distance & Variance of feature cluster centers, generated from a multivariate Normal & [0.0,2.0]\\
    feature dim & Dimension of feature vector assigned to every node & [16,16]\\
  \bottomrule
\end{tabular}
\end{table*}

\begin{table*}
  \caption{LFR}
  \label{tab:LFR}
  \begin{tabular}{lll}
    \toprule
    Parameter Name & Description & Values\\
    \midrule
    nvertex & Number of vertices & [1028,4096] \\
    avg. degree & Average node degree & [1,32]\\
    max degree proportion & Minimum degree, as a proportion of the node count & [2,20] \\
    mixing param & Ratio of total number of inter to intra community edges  & [0.0,1.0]\\
    min community size proportion & Minimum size of community, as a proportion of the node count & [0.05, 0.0825]\\
    max community size proportion & Maximum size of community, as a proportion of the node count & [0.25,0.33]\\
    community exponent & Value of the power law exponent used to generate community sizes & [1.0,2.0]\\
    exponent & Value of the power law exponent used to generate node degrees & [2.0,3.0]\\
    num tries & Number of attempts at simulating an LFR graph until success & [20,20] \\
    feature center distance & Variance of feature cluster centers, generated from a multivariate Normal & [0.0,2.0]\\
    feature dim & Dimension of feature vector assigned to every node & [16,16]\\
  \bottomrule
\end{tabular}
\end{table*}

\end{document}